%% file: sample-sigconf.tex
\documentclass[sigconf]{acmart}
\usepackage{booktabs} % For formal tables
\usepackage{enumitem}
\usepackage{multirow}
\usepackage[utf8]{inputenc}
\makeatletter
\renewcommand\@formatdoi[1]{\ignorespaces}
\makeatother
\setcopyright{rightsretained}
\acmDOI{}
\acmISBN{}
\acmConference[RecSysKTL'18]{Workshop on Intelligent Recommender Systems by Knowledge Transfer and Learning}{October 2018} {Vancouver, Canada} 
\acmYear{2018}
\copyrightyear{2018}
\acmArticle{}
\acmPrice{}

\newcommand {\otoprule }{\midrule[\heavyrulewidth]}

\begin{document}
\title{Do Better ImageNet Models Transfer Better... for Image Recommendation?}

%\author{Author et al}
%\affiliation{%
%  \institution{Monster University}
%  \streetaddress{A Street}
%  \city{A City} 
%  \state{A Country} 
%}
%\email{author1@monster.edu}

\author{Felipe del Rio} 
\affiliation{%
  \institution{IMFD \& PUC Chile}
  \city{Santiago} 
  \state{Chile} 
}
\email{fidelrio@uc.cl}

\author{Pablo Messina} 
\affiliation{%
  \institution{IMFD \& PUC Chile}
  \city{Santiago} 
  \state{Chile} 
}
\email{pamessina@uc.cl}

\author{Vicente Dominguez} 
\affiliation{%
  \institution{IMFD \& PUC Chile}
  \city{Santiago} 
  \state{Chile} 
}
\email{vidominguez@uc.cl}

\author{Denis Parra} 
\affiliation{%
  \institution{IMFD \& PUC Chile}
  \city{Santiago} 
  \state{Chile} 
}
\email{dparra@ing.puc.cl}

\renewcommand{\shortauthors}{del Rio et al.}

\begin{abstract}
% context
Visual embeddings from Convolutional Neural Networks (CNNs) trained on the ImageNet dataset for the ImageNet Large Scale Visual Recognition Competition (ILSVRC) challenge have shown consistently good performance for transfer learning and are  widely used in several tasks, including image recommendation. 
% our aim
However, some important questions have not yet been answered in order to use these embeddings for a larger scope of recommendation domains: a) Do CNNs that perform better in ImageNet also work better for transfer learning in content-based image recommendation?, b) Does fine-tuning help to improve performance? and c) Which is the best way to perform the fine-tuning?

% what we do
In this paper we compare several CNN models pre-trained with the ImageNet dataset to evaluate their transfer learning performance to an artwork image recommendation task. Our results indicate that models with better performance in the ImageNet challenge do not always imply better transfer learning for artistic image recommendation tasks (e.g., NASNet vs. ResNet). Further analysis shows that fine-tuning can be helpful even with a small dataset, but not every fine tuning works.

%implications
Our results, although preliminary and focused on the art domain, can inform other researchers and practitioners on how to train their CNNs for better transfer learning towards image recommendation systems.

\end{abstract}



\vspace{-2mm}

\maketitle

\input{samplebody-conf}

\vspace{-2mm}
\bibliographystyle{ACM-Reference-Format}
\bibliography{sigproc} 

\end{document}

%% file: samplebody-conf.tex
\vspace{-2mm}
\section{Introduction}
% Context, importance of the problem
%Despite the large financial crisis started in 2008, the online artwork market has kept growing over time. 
%Compared to markets affected by 2008's financial crisis, online artwork sales are booming due to social media and new consumption behavior of millennials. Online art sales reached \$3.27 billions in 2015, and at the current grow rate, they will reach \$9.58 billion by 2020 \cite{forbes1}. 
% lack of recsys compared to other domains
%Notably, although many online businesses utilize recommendation systems to boost their revenue, online artwork recommendation has received little attention compared to other areas such as movies \cite{amatriain2013mining} or music \cite{celma2010music}. 
%Previous research

The outstanding results of Deep Convolutional Neural Networks (CNN) models in the area of computer vision since 2012 \cite{krizhevsky2012imagenet} as well as their performance for transfer learning to different datasets and tasks such as medical image classification \cite{hoo2016deep} and image classification in small datasets \cite{oquab2014learning} have made these models an important component in areas such as image-based recommendation.
Several works in recommender systems \cite{he2016vbpr,he2016vista,dominguez2017,messina2018} have used CNNs such as AlexNet \cite{krizhevsky2012imagenet} or VGG \cite{simonyan2014vgg19} to automatically extract the features representing an image as a vector of visual features. This embedding is eventually used to train other models \cite{he2016vbpr,he2016vista} or to directly match and recommend similar images \cite{dominguez2017,messina2018}. However, an implicit assumption about these models is that the better they perform in the ImageNet Large Scale Visual Recognition Competition (ILSVRC) \cite{ILSVRC15} in the ImageNet dataset, the better they will perform in other tasks. Kornblith et al. \cite{kornblith2018imagenetmodelstransfer} challenged this assumption by evaluating the capacity of several state-of-the-art CNN models for transfer learning to different computer vision datasets. They showed that there is a rather small correlation between ImageNet performance and transfer learning performance when these CNNs are used solely as fixed feature extractors. All the models improved though after fine-tuning. The results provided important insights about using these CNNs models for transfer learning: always using the top performing model in the ILSVRC as a pre-trained visual feature extractor is not always the best idea. However, they did not test CNN visual transfer learning in an image recommendation task. 

\paragraph{\textbf{Objective}} In this article, motivated by the experiments and results by Kornblith et al. \cite{kornblith2018imagenetmodelstransfer}, we study transfer learning for image recommendations. More particularly, we investigate whether the performance of a CNN model in the ImageNet dataset correlates with the results of recommending artwork images. Moreover, we experiment with two different fine-tuning alternatives (deep and shallow) to find out if we can improve the performance of the CNN when used solely as a feature extractor.

%Previous research has shown the potential of personalized recommendations in the arts domain, such as the CHIP project \cite{aroyo2007personalized}, that implemented a personalized recommendation system for the Rijksmuseum. More recently, He et al. \cite{he2016vista} used pre-trained deep neural networks (CNN) for recommendation of digital art, obtaining good results. Unfortunaly, their method is not applicable for the physical artwork problem as the method assumes that the same item can be bought over and over again. Hence their work only works under the collaborative filtering assumption and also did not investigate explicit visual features nor metadata.

%	Objective
%\textbf{Objective}. In this paper, we investigate the impact of different features for recommending physical artworks. In particular, we reveal the utility of artwork metadata, latent (CNN) and explicit visual features extracted from images. We address the problem of artwork recommendation with  positive-only feedback (user transactions) over \emph{one-of-a-kind} items, i.e., only one instance of each artwork (paintings) is available in the dataset.

% Our research questions

\paragraph{\textbf{Research Questions}} Our work was driven by the following research questions: 

\begin{itemize}[leftmargin=*]

\item \emph{RQ1}. Are CNNs that perform better in ImageNet also better for transfer learning in content-based image recommendation?
\item \emph{RQ2}. Does fine-tuning help to improve the performance in the image recommendation task? 
\item \emph{RQ3}. What  is the best way to perform the fine-tuning?

\end{itemize}

\begin{figure}[t!]
    \includegraphics[width=\linewidth,trim={1.1cm 0.9cm 1.4cm 0.5cm},clip]{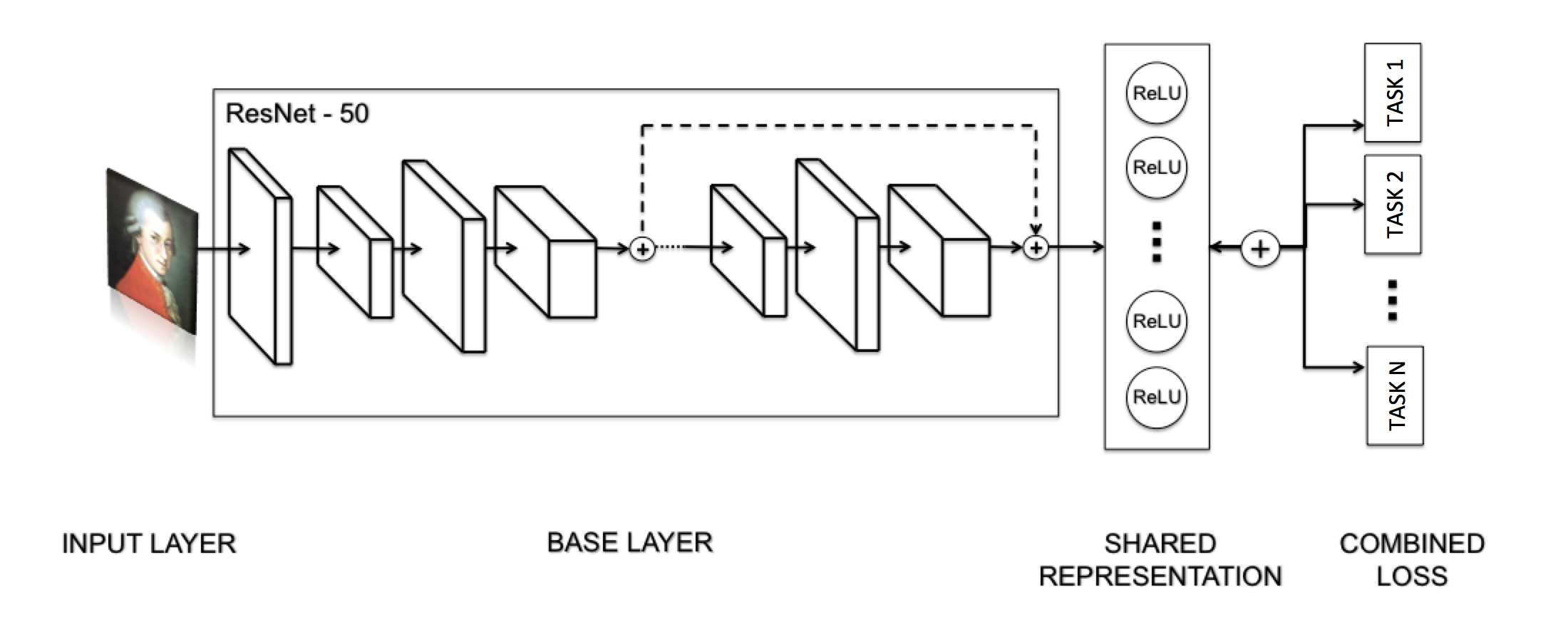}
    \caption{Generic model used for fine-tuning, adapted from Strezoski et al. \cite{strezoski2017omniart}. In the case of {\it shallow tuning} only the Shared Representation layer is updated. On the other side, for {\it deep tuning}, all the convolutional layers are tuned.}
    \label{fig:model}
\vspace{-7mm}    
\end{figure}

\paragraph{\textbf{Contributions}} Our work contributes to the area of transfer learning for image and image-based recommendation. We experiment with five pre-trained state-of-the-art CNN models for the ILSVRC competition. We also run simulated experiments to predict artwork purchases using a real-world transaction data provided by a popular online artwork store based in USA named \textit{UGallery}\footnote{\url{www.ugallery.com}}.  We also show how to fine-tune the best of these pre-trained models in order to boost its performance for a recommendation task. Our results are at the level of the state-of-the-art on content-based artwork recommendation \cite{messina2018}.

\vspace{-2mm}
\section{Datasets} 

%\vspace{-2mm}
\subsection{UGallery} 
UGallery provided us with an anonymized dataset of 2,919 users, 6,040 items and 4,099 purchases (transactions) of paintings, where all users made at least one transaction. On average, each user has bought 2-3 items %$2.08\pm2.89$  items
in the latest years\footnote{Our collaborators at UGallery asked us not to disclose the exact dates when the data was collected.}. Each painting in this dataset is unique, so once an artwork is bought it is not available anymore to be recommended or to calculate recommendations based on co-occurrences, such as collaborative filtering \cite{resnick1994grouplens,parra2013recommender}.

%The distribution is skewed since most users (871 in total) bought only one item, and only a few users (53 in total) have bought 7 or more items. Our data is not atypical, since it resembles the rating distribution on the Netflix prize or the Movielens dataset, where a few users account for most of the activity and most users have little or none \cite{Harper2015,bennett2007netflix}. 

\textbf{Metadata}. Artworks in the \textit{UGallery} dataset were manually curated by experts. In total, there are five attributes, but only two of these attributes are present in all paintings: medium (e.g., oil, acrylic), and artist. We use these attributes to fine-tune the pre-trained models described later in Section \ref{sec:methods}.

\textbf{Visual Features}. For each image representing a painting in the dataset we obtain features from CNN models described later in the Section \ref{sec:methods}.

\vspace{-2mm}
\subsection{Omniart} 

For some fine-tuning experiments we used the large art dataset Omniart \cite{strezoski2017omniart}, which originally reported of containing 432,217 artwork images, but has since grown to include more than 1 million data samples. Each data sample contains an image and metadata associated to it. These metadata include \textit{artwork name}, \textit{artist full name}, \textit{year of creation}, \textit{collection origins}, \textit{general type}, \textit{artwork type}, \textit{school} and \textit{number of creators}, among others.

In order to use this dataset for fine-tuning, we cleaned it, removing fields that were mostly empty. We then kept 3 fields: artist, artwork type and year of creation. Next, we filtered out every sample with at least one of the aforementioned fields empty. We also filtered out all the artworks by artists that appeared less than 100 times and we did the same for the artwork type. We ended up with 634,508 images, containing 2,080 different artists and 47 different artwork types. This dataset was then divided into train, validation and test sets in a proportion of 70\%, 20\% and 10\% respectively.

\vspace{-2mm}
\section{Methods}
\label{sec:methods}

%\vspace{-2mm}
\subsection{Transfer Learning}
\label{sec:transfer-learning}

We experimented using some of the top performing CNNs models for the ImageNet challenge as presented in the updated list of TensorFlow-Slim, a high-level Tensorflow API for image classification \cite{tensorflow-slim2016}. The models selected for comparison in this article were \textbf{VGG19}~\cite{simonyan2014vgg19}, \textbf{ResNet50}~\cite{he2015resnet}, \textbf{InceptionV3}~\cite{szegedy2015inceptionv3}, \textbf{InceptionResNetV2}~\cite{szegedy2016inceptionresnet} and \textbf{NASNet}~\cite{zoph2017nasnet}.

As pointed out by Kornblith et al. \cite{kornblith2018imagenetmodelstransfer}, it is commonly assumed that the best performing CNN models in the ILSVRC challenge are also going to be the top performing models in other visual tasks. Following this assumption, we should expect  \textbf{NASNet} to output the best embeddings for image recommendation. In the same work, Kornblith et al. \cite{kornblith2018imagenetmodelstransfer} found that there is no perfect correlation between the performance of a model in the ILSVRC and in other visual tasks, as ResNet performs better than other models when using the pre-trained features obtained from it. We would like to find out if this result remains in our task of recommending art.

% ============================
% Results table for transfer learning
% ============================

\newcommand{\ra}[1]{\renewcommand{\arraystretch}{#1}}
\begin{table}[!t]
\centering
\caption{Results of different pre-trained embeddings at the artwork image recommendation task to the left (R:Recall, P:Precision), and their performance at the ILSVRC Challenge trained on ImageNet dataset (Acc: Accuracy). The top methods in both tasks do not correlate.}
\vspace{-3mm}
\label{tab:tf-results}
\scalebox{0.62}{
    \setlength{\tabcolsep}{1.2em} % for the horizontal padding
    \ra{1.3} % for the vertical padding
    \begin{tabular}{@{\extracolsep{-10pt}}lcccccc}
        \otoprule
        % headers
        \multicolumn{1}{c}{\multirow{2}{*}{\textbf{CNN}}} &
        \multicolumn{4}{c}{Artwork Image Recommendation} &
        \multicolumn{2}{c}{ILSVRC-2012-CLS}\\
        \cmidrule(l{9pt}r{9pt}){2-5}
        \cmidrule(l{9pt}r{9pt}){6-7}
        % rest
        \multicolumn{1}{c}{} & \textbf{R@20} & \textbf{P@20} & \textbf{MRR@20} & \textbf{nDCG@20}
        & \textbf{Top-1 Acc. (\%)} & \textbf{Top-5 Acc. (\%)} \\

        \otoprule
        ResNet50          &    \textbf{.1632} &     \textbf{.0141} &    \textbf{.0979} &     \textbf{.1253} & 75.2 & 92.2 \\
        VGG19             &    \textbf{.1398} &     \textbf{.0124} &    \textbf{.0750} &     \textbf{.1008} & 71.1 & 89.8 \\
        NASNet Large      &    .1379 &     .0120 &    .0743 &     .0998 & \textbf{82.7} & \textbf{96.2} \\
        InceptionV3       &    .1332 &     .0125 &    .0744 &     .1007 & 78.0 & 93.9 \\
        InceptionResNetV2 &    .1302 &     .0117 &    .0692 &     .0936 & \textbf{80.4} & \textbf{95.3} \\
        Random                  &    .0172 &     .0013 &    .0051 &     .0093 & - & - \\
        \bottomrule
    \end{tabular}
}
\vspace{-5mm}
\end{table}
% ============================
% Results table for transfer learning
% ============================

% ============================
% Results table for fine-tuning
% ============================

\begin{table*}[!h]
\centering
\caption{Results of the simulated recommendation experiment. Notice how a shallow fine-tuning of the ResNet model with the Omniart dataset decreases the performance of the model, while a deep fine-tuning of all layers with the small UGallery dataset improves performance of ResNet and even further for the Omniart model.}
\vspace{-2mm}
\label{tab:results-finetuning}
\scalebox{0.78}{
\begin{tabular}{lcccccc}
\otoprule
%  --- header
\textbf{CNN} & \textbf{R@20} & \textbf{P@20} & \textbf{F1@20} &
\textbf{MAP@20} & \textbf{MRR@20} & \textbf{nDCG@20} \\
[0.5ex]\midrule
% ---- body
 %CNN (Omniart-fine-tune-all-layers)        &    .1954 &     .0164 &     .0276 &    .0294 &    .1155 &     .1476 \\
ResNet-deep-fine-tune-ugallery        &    .1954 &     .0164 &     .0276 &    .0294 &    .1155 &     .1476 \\
ResNet-deep-fine-tune-ugallery-only-artist               &    .1943 &     .0166 &     .0279 & .0300 &    .1166 &    .1493  \\
Omniart-deep-fine-tune-ugallery               &    .1900 &     .0159 &     .0266 & .0267 &    .0973 &    .1330  \\
[0.5ex]\midrule
ResNet    &    .1632 &     .0141 &   .0235 & .0246 & .0979    &    .1253 \\
[0.5ex]\midrule
Omniart-shallow-with-task-weights        &    .1609 &     .0134 &     .0224 &    .0227 &    .0879 &     .1147 \\
 
ResNet-shallow-fine-tune-ugallery-only-artist               &    .1501 &     .0137 &     .0230 &    .0242 &    .0936 &     .1202 \\
ResNet-shallow-fine-tune-ugallery               &    .1541 &     .0138 &     .0229 &    .0238 &    .0942 &     .1196 \\
ResNet-shallow-fine-tune-ugallery-only-medium               &    .1541 &     .0138 &     .0225 &    .0238 &    .0894 &     .1165 \\
  
Omniart-shallow-only-type               &    .1510 &     .0127 &     .0212 &    .0217 &    .0831 &     .1092 \\
Omniart-shallow-no-task-weights                    &    .1473 &     .0129 &     .0214 &    .0234 &    .0906 &     .1150 \\
Omniart-shallow-only-artist             &    .1442 &     .0129 &     .0213 &    .0235 &    .0908 &     .1153 \\
ResNet-deep-fine-tune-ugallery-only-medium              &    .1374 &     .0124 &     .0204 &    .0218 &    .0856 &     .1101 \\

Omniart-shallow-only-period             &    .0937 &     .0081 &     .0135 &    .0127 &    .0514 &     .0689 \\
Random                                   &    .0172 &     .0013 &     .0022 &    .0014 &    .0051 &     .0093 \\
\bottomrule
\end{tabular}
}

\vspace{-3mm}
\end{table*}
% ============================
% Results table for fine-tuning
% ============================

\vspace{-3mm}
\subsection{Fine-tuning}

% We experimented with two different models of feature extraction from the artwork images in order to make content-based recommendations.
Beyond using each CNN solely as a pre-trained model for visual feature extraction and eventual transfer learning, we experimented with two different processes for fine-tuning the CNNs with the aim of improving their performance in artwork recommendation: shallow and deep fine-tuning. In addition, the fine-tuning could be performed by approaching single-task or multitask learning.

%Based on the evidence showing that features extracted from convolutional neural networks (CNNs) are effective in art recommendation \cite{he2016vista,dominguez2017,messina2018} and on the model proposed by Strezoski et al. \cite{strezoski2017omniart} for learning an art model to perform multitask data analysis, we propose a deep convolutional neural network that leverages multitask learning to extract features from artwork images. The model is flexible, and the same network can be adapted for single-task learning if necessary.

Our fine-tuning method is based on the work by Strezoski et al. \cite{strezoski2017omniart}, who propose a deep CNN that leverages multitask learning to extract features from artwork images. The model is flexible, and the same network can be adapted for single-task learning.
As seen in Figure \ref{fig:model}, our generic network model for fine-tuning is composed of three different parts which are stacked in the same order as presented. First, a base representation layer (Base Layer) acts as feature extractor. Our work is agnostic to the choice of this layer, but for the example we used ResNet50. Next, add a densely connected shared representation layer, of size $1,024$. This layer is then connected to one-to-many output layers, one for each task used to train the network. The shared representation layer allows the model to learn a rich representation that is useful to each of the tasks learned. After fine-tuning, we will extract the image visual features used for recommendation based on the activation of the shared representation layer.

To train the model, we stage a supervised task based on one of our two datasets. This supervised task consists in predicting the artist, type, medium, etc. labeled in for the artwork image. Each output layer focuses on one task, and the loss that is used depends on the task type: Cross Entropy for classification tasks (such as artist  and type prediction), and MAE for regression tasks (such as period in the OmniArt dataset \cite{strezoski2017omniart}).

%\paragraph{\textbf{Shallow Fine-tuning}}
\textbf{Shallow Fine-tuning}. In this fine-tuning method we keep the base layer frozen at training time, so we only adjust the weights of the {\emph shared representation layer}. Thus, we use the output of the pre-trained network as features of the image, without re-training it.

%\paragraph{\textbf{Deep Fine-tuning}}
\textbf{Deep Fine-tuning}. Based on the insight given by Kornblith et al. \cite{kornblith2018imagenetmodelstransfer} as well as previous works on the medical image domain \cite{tajbakhsh2016convolutional}, we update all the weights in the base layer, shown in Figure \ref{fig:model}. The aforementioned works indicate that a deep fine-tuning usually increases a model performance when transferring to other classification tasks, and even with a small dataset the improvement can be significant.
%and training a regression, we tested our previous model but instead of just training the last layer, we trained the whole network.

%\paragraph{\textbf{Single task vs. multitask learning}}
\textbf{Single task vs. multitask learning}. We experimented using multiple tasks. The argument for pursuing multitask learning lies in the expectation of learning more flexible embedding than single-task learning, since the same representation must be useful to solve several tasks. Then, for our multitask learning approaches, our loss functions are inspired by a simplified version of the one used by Strezoski et al. \cite{strezoski2017omniart}. Being $L$ the cumulative loss for all tasks, $L_i$ the loss for task $i$ and $w_i$ the weight associated to task $i$, as a way to increase the importance of any given task, the multitask loss is: %\ref{eq:loss-eq}. 
\vspace{-2mm}

\begin{equation}
\vspace{-1mm}
    L=\sum_{i=0}^{N}{w_i*L_i}
    \label{eq:loss-eq}
\vspace{-1mm}
\end{equation}

\subsection{Training}

We used an Adam optimizer for all of our training with a learning rate of $0.001$ when training on the OmniArt Dataset, and $0.0001$ when training on the UGallery dataset as this increases the performance on the prediction task. We used a validation set in order to avoid overfitting, and when the validation error did not decrease in 5 epochs we considered the training as complete.

\vspace{-2mm}
\section{Recommendation Task: Predict purchases}

Our experimental protocol for evaluating image recommendation performance is based on our previous work \cite{dominguez2017,messina2018}. We predicted the items bought in each transaction of the dataset using the previous purchases of a customer to build a user model. We then provide recommendations by matching images in the UGallery dataset with the largest cosine similarity to the images in the user profile.  

\vspace{-2mm}
\section{Results}

We first present the results of using pre-trained models for static visual feature extraction, in Table \ref{tab:tf-results}. Afterwards, we select the best performing model and proceed with the results about fine-tuning this model for image recommendation, in Table \ref{tab:results-finetuning}.

\vspace{-2mm}
\subsection{RQ1: Transfer Learning}

Consistent with the results of Kornblith et al. \cite{kornblith2018imagenetmodelstransfer}, we find that there is no correlation between performance in ILSVRC trained with ImageNet and the actual image recommendation task, as clearly seen in Table \ref{tab:tf-results}. The top methods in ILSVRC are NASNet and InceptionResNetV2, but in the artwork image recommendation task the top performing methods were ResNet and VGG19. ResNet also was the best pre-trained visual feature extractor in the series of experiments conducted by Kornblith et al. \cite{kornblith2018imagenetmodelstransfer}, which indicates the quality of this network embedding to transfer learning across different datasets and tasks. Subsequently, we use ResNet for the fine-tuning task presented next.

\vspace{-2mm}
\subsection{RQ2 \& RQ3: Fine-Tuning}

Table \ref{tab:results-finetuning} presents the results of our fine-tuning and multitask experiments.
% 1. influencia del dataset: fine tuning con Ugallery, que es el dataset usado en la tarea de recomendacion, reportó mejor resultado
We find that there is a significant influence on the dataset used, when using the actual dataset from which recommendations are going to be drawn. The performance of the method increased substantially evidenced by the top-3 best performances are achieved when fine-tuning with UGallery dataset.

% 2. fine-tuning con Omniart: no funcionó bien, bajó el rendimiento respecto de ResNet
Surprisingly, the use of a larger dataset of artwork images with metadata, such as Omniart, does not help in the recommendation task, and even decreases the performance if not fine-tuned with UGallery. This might be explained because of important differences between the samples from both the training dataset (Omniart) and the recommendation dataset (UGallery). For instance, the latter contains more modern and abstract artworks, and the former contains pictures, masks, sculptures, and pottery. Filtering out types other than paintings remains for a future work to explore.

% 3. Tipo de fine-tuning: deep fine tuning funcionó mucho mejor
Also consistent with the results of Kornblith et al. \cite{kornblith2018imagenetmodelstransfer}, we found that the performance of a shallow fine-tuning is significantly lower than the performance achieved when using deep fine-tuning. All of the top performing methods are achieved using this fine-tuning variant.

% 4. Multitask vs. single-task : no hay diferencias grandes entre multitask y single-task con artist, pero sí entre multitask y single-task period
Moreover, when \textit{artist} was the target of the learning task, we did not find significant differences between multitask and single-task learning. However, we identified a significant difference  when the learning task includes the targets \textit{period} or \textit{medium}. The latter achieved a lower performance most of the time. This result could be explained by the distributions of \textit{mediums types} in the UGallery dataset, since more than 60\% of the artworks belong to the medium type \textit{oil painting}. Hence, learning this classification might provide little information about personal user preference.

\vspace{-2mm}
\section{Conclusions}

In this paper, motivated by the previous works of Kornblith et al. \cite{kornblith2018imagenetmodelstransfer} in transfer learning, as well as Strezoski and Worrying \cite{strezoski2017omniart} research on multitask learning for art images, we have studied whether the accuracy of several top-performing pre-trained models in the ImageNet dataset correlates with their transfer learning capacities for image recommendation. Our results indicate that there is no clear correlation, and that a neural model like ResNet performs better for transfer learning to image recommendation compared to NASNet of InceptionResNetV2, which performs better in the ILSVRC task. 

Using ResNet as our base model, we also tested several fine-tuning alternatives to improve the performance of the pre-trained CNN model. We found that a deep fine-tuning (rather than a shallow fine-tune) along with the same target dataset UGallery (rather than the larger but noisy Omniart), can significantly improve the performance of the recommendations evaluated in several metrics. With respect to multitask vs. single-task learning, there was no a clear winner, but results seem to indicate that learning an embedding that discriminates between artists can be helpful for predicting artistic image preferences.

Our results are good, improving our best performance in this task in \cite{dominguez2017}, and have the advantage of not relying on the metadata of the artowork, just on the image. This combined with the current tooling that has been developed \cite{chollet2015}\cite{paszke2017automatic} for the use of deep models makes this approach very convenient.

In future work we expect to test the fine-tuning performance of all the pre-trained models (not only ResNet). Moreover, we will test other datasets in an attempt to generalize our results to other domains of image recommendation beyond art. Finally, we will attempt to test different neural network architectures to learn user preferences, such as a siamese network or one supporting the triplet loss.

\vspace{-2mm}
\section{acknowledgements}
The authors from PUC Chile were funded by Conicyt, Fondecyt grant 11150783, as well as by the Millennium Institute for Foundational Research on Data (IMFD).